\documentclass[journal,10pt]{IEEEtran}

\usepackage{cite}
\usepackage{epsfig}
\usepackage{epstopdf}
\usepackage{graphicx}
\usepackage[dvipsnames]{xcolor}
\usepackage{tikz}
\usetikzlibrary{arrows,positioning,shapes.geometric,circuits.logic.US}
\usepackage{pgfplots}
\usepackage{multirow}
\usepackage{upgreek}
\usepackage{amssymb}
\usepackage{amsmath}
\usepackage{cases}
\usepackage{url}
\usepackage{pifont}
\usepackage{bm}
\usepackage{amsthm}

\usepackage{tabularx}
\usepackage{booktabs}
\usepackage{filecontents}
\usepackage{subcaption}
\newcolumntype{Y}{>{\centering\arraybackslash}X}

\theoremstyle{definition} 
\theoremstyle{definition} 
\theoremstyle{definition} 
\theoremstyle{definition}

\usepackage[linesnumbered, ruled, vlined, noend]{algorithm2e}

\usepackage{algorithmic}

\usepackage{array}

\newcommand{\fixme}[2]{\ifx&#2&{\leavevmode\color{red}#1}\else{\leavevmode\color{red}FIXME\{}#1{\leavevmode\color{red}\}}\footnote{{\leavevmode\color{red}#2}}\PackageWarning{Fixme}{#1: #2}\fi}

\newcommand{\newstuff}[2]{\ifx&#2&{\leavevmode\color{blue}#1}\else{\leavevmode\color{blue}FIXME\{}#1{\leavevmode\color{blue}\}}\footnote{{\leavevmode\color{blue}#2}}\PackageWarning{Newstuff}{#1: #2}\fi}

\title{Input-distribution-aware successive cancellation list decoding of polar codes}

\author{\IEEEauthorblockN{Carlo~Condo,~\IEEEmembership{Senior~Member,~IEEE}\\}
\IEEEauthorblockA{Infinera Corporation\\
Email: ccondo@infinera.com}} 

\begin{document}

\maketitle
\begin{abstract}
Polar codes are linear block codes that can achieve channel capacity at infinite code length. 
Successive cancellation list (SCL) decoding relies on a set of parallel decoders; it yields good error-correction performance at finite code length, at the cost of increased implementation complexity and power consumption. 
Current efforts in literature focus on design-time decoder complexity reduction, while lacking practical run-time power reduction methods.
In this work, input-distribution-aware SCL (IDA-SCL) decoding is proposed, that allows to determine the parallelism to adopt by performing simple observations on the input of the decoder.  
This technique guarantees fixed, short latency and allows hardware SCL decoders to dynamically shut down part of the internal parallelism before each decoding process. 
It can be combined with existing complexity- and power- reduction techniques. 
Simulation results show that IDA-SCL can reduce the run-time complexity of SCL of up to 50\%.

\end{abstract}


\IEEEpeerreviewmaketitle

\section{Introduction} \label{sec:intro}

Polar codes \cite{ArikanFirst} are linear block codes that are able to achieve the capacity of binary-input memoryless channels at infinite code length. 
They have been the subject of growing attention and research efforts in the academic and industrial research community, that have striven to improve their finite-length error correction performance, reduce their decoding latency, and design low-complexity, low-power decoder implementations. 
Thanks to the outstanding results obtained in the last decade, polar codes have been included in the 3GPP 5th generation wireless systems standard (5G) \cite{3GPP_R15}, and have made their way towards optical communications \cite{Mehmood:20,PPCjournal}. 

Successive cancellation list (SCL) decoding has been proposed in \cite{TalSCL} to improve on the error-correction performance of polar codes provided by their original decoding algorithm, successive cancellation (SC). 
It relies on a list of parallel SC decoders, each making different decoding choices; with its countless evolutions and ameliorations, it is considered the academic and industrial standard for polar code decoding. 
Unfortunately, the improved performance of SCL decoding comes at an increased implementation cost.
Various techniques are available in literature to reduce the complexity and power consumption of SCL decoders. 
Design-time approaches like \cite{PartSCL,PartSCL_TCOM, ANNpredictor} modify the structure of the basic SCL decoder, reducing its implementation complexity. 
However, at the latest technology nodes, dynamic power dominates the total power consumption: it is thus of paramount importance to combine efficient design to run-time power-reduction techniques. 
The adaptive SCL (ASCL) decoder described in \cite{ASCL} can potentially reduce the average power consumption by performing sequential decoding attempts with increasing list size. 
However, it introduces variable decoding latency, an undesired feature in any practical decoder implementation, that needs to be timed according to the worst case latency. 
To use ASCL under fixed latency, the system has to accept a decoding delay equal to the duration of all possible sequential decoding attempts.

In this work, input-distribution-aware SCL (IDA-SCL) is proposed, a technique that allows to decide which list size to adopt by observing the input of the decoder.  
This technique guarantees fixed, short decoding latency and allows hardware SCL decoders to dynamically decrease the list size before each decoding by shutting down part of the internal parallelism. 
It relies on simple operations that have negligible implementation cost, and it can be stacked with any other complexity and power reduction technique. 
Simulation results over a wide set of code parameters show that IDA-SCL can reduce the run-time complexity of SCL of up to 50\%.

\section{Preliminaries} \label{sec:prel}

A polar code of length $N$ relies on a transformation matrix $T_N = T_2^{\otimes n}$, generated by the $n$-fold Kronecker product of a basic channel transformation kernel $T_2$. 
The resulting $N$ bit-channels are polarized, and vary from completely noisy to completely noiseless. 
A polar code of length $N$ and rate $R=K/N$ is constructed by creating an input vector $\boldsymbol{u}$ where the $K$ message bits are assigned to the entries of $\boldsymbol{u}$ corresponding to the $K$ most reliable bit-channels.
The remaining entries of $\boldsymbol{u}$ are ``frozen" bits, set to zero. 
The codeword $\boldsymbol{x}$ is then calculated as $\boldsymbol{x} = \boldsymbol{u} \cdot T_N$ and transmitted.

\begin{table*}[t!]
  \caption{Percentage of cases for $L_{low}$ requirement, $L_{high}=32$.}
	\centering
	\begin{tabular}{|c|c|c|c|c|c|c|}
		\hline
		\multirow{2}{*}{$L_{low}$} & \multicolumn{2}{c|}{$N=128$} & \multicolumn{2}{c|}{$N=256$} & \multicolumn{2}{c|}{$N=512$} \\
		\cline{2-7}
		    &	$R=1/2$ $C=2$ &	 $R=7/8$ $C=3$ & $R=1/2$ $C=5$ & $R=7/8$ $C=8$ & $R=1/2$ $C=8$ & $R=7/8$ $C=12$ \\
   		\hline
		1  	 &	96.87\%		&	98.60\%		&		84.61\%	&	94.04\% & 84.87\%	&	88.78\%		\\
		2	&	2.52\%		&	1.10\%		&		11.44\%	&	3.96\%	&	10.67\%	&	7.89\%		\\
		4	&	0.45\%		&	0.23\%		&		2.85\%	&	1.42\%	&	2.99\%	&	2.28\%		\\
		8	&	0.12\%		&	0.05\%		&		0.76\%	&	0.40\%	&	1.00\%	&	0.65\%		\\
		16	&	0.03\%		&	0.02\%		&		0.23\%	&	0.15\%	&	0.34\%	&	0.28\%		\\
		32	&	0.01\%		&	$<$0.01\%		&		0.11\%	&  0.03\%	  	&	0.13\%	&		0.12\%	\\
		\hline	
	\end{tabular}
	\newline\newline
\label{tab:minL}
\end{table*}


SC decoding was proposed in \cite{ArikanFirst}.
It can be represented as a depth-fist binary tree search with priority to the left branch. 
The logarithmic likelihood ratio (LLR) vector $\boldsymbol{y}$ is received from the channel and assigned to the root node: the LLRs are propagated, through node operations, downward towards the leaf nodes, each associated to an entry of the estimated input vector $\hat{\boldsymbol{u}}$. 
Bit values at leaf nodes are either known (in case of frozen bits) or estimated based on the sign of the incoming LLR. 
Bit estimations are propagated upward and combined with the descending LLRs until all leaves have been explored.

While it can achieve capacity at infinite code length, SC decoding has mediocre error-correction performance at moderate code length. 
To improve it, a list-based decoder has been proposed in \cite{TalSCL}. 
The SCL decoder relies on $L$ parallel SC decoders, each one storing a different partial decoded bits vector, called \textit{path}. 
Every time an information leaf node is reached, each decoder splits the current decoding path, estimating the bit as $0$ in one case and as $1$ in the other, doubling the number of parallel decoding paths. 
A path metric allows to maintain only the $L$ more likely paths, while the $L$ less likely are discarded. 
The decoding continues until the last leaf node has been reached, and one as the surviving decoding paths is selected as the decoder output. 
The concatenation of polar codes with a cyclic redundancy check (CRC) of length $C$ has been proposed in \cite{CA_SCL} to help the final decoder output path selection, showing substantial improvement. In the remainder of the paper, this is the considered version of SCL.

\section{input-distribution-aware SCL decoding}\label{sec:MESCL}

Given a polar code of code length $N$ and rate $R$, and a codeword transmitted through an additive white Gaussian noise (AWGN) channel characterized by a certain $E_b/N_0$, every received vector that can be successfully decoded through SCL with $L=L_{high}$, may be decoded by SCL with $L=L_{low}\le L_{high}$. 
Table \ref{tab:minL} reports the percentage of cases for which a particular $L_{low}$ is the minimum power of 2 necessary for the decoder to select the correct codeword as the output, for various combinations of $N$, $R$ and $C$, at a block error rate (BLER) of approximately $10^{-3}$, and $L_{high}=32$. 
Such a BLER is a realistic working point for many wireless applications, and for component codes in more powerful concatenated coding schemes targeting optical communications.
Simulation results consider $10^6$ frames.  
It can be seen that the vast majority of cases does not need $L=32$ to be correctly decoded, with the percentage rising as $N$ decreases and $R$ increases. 
This is also due to the fact that the required list size strongly depends on the input noise level, and that given a BLER target, different combinations of code parameters achieve it at different $E_b/N_0$. 

\begin{figure}[t!]
  \centering
  \begin{tikzpicture}
  \pgfplotsset{
    label style = {font=\fontsize{8pt}{7.2}\selectfont},
    tick label style = {font=\fontsize{6pt}{7.2}\selectfont}
  }

\begin{axis}[
	scale = 1,
    xlabel={LLR value}, xlabel style={yshift=-0.2em},
    ylabel={Average number of LLRs}, ylabel style={yshift=-0.75em},
    grid=both,
    ymajorgrids=true,
    xmajorgrids=true,
    grid style=dashed,
    mark options=solid,
    ybar interval,
    ymax = 256,
    ymin = 0,
    width=1\columnwidth, height=5.8cm,
    thick,
	xmin=1,
	xmax=12,
	xtick={1,2,3,4,5,6,7,8,9,10,11,12},
    xticklabels={$<$0.5,$<$1.0,$<$1.5,$<$2.0,$<$2.5,$<$3.0,$<$3.5,$<$4.0,$<$4.5,$<$5.0,$<$5.5,$<$6.0},
    legend style={
      anchor={center},
      cells={anchor=west},
      mark options=solid,
      column sep= 2mm,
      font=\fontsize{7pt}{7.2}\selectfont,
    },
    legend to name=LLRdist,
    legend columns=3,
]

\addplot[
	color=CornflowerBlue,
	fill,
]
table {
1 16.13	
2 32.708	
3 50.033	
4 68.297	
5 87.462	
6 107.3	
7 127.37
8 147.11	
9 165.96
10 183.39
11 198.97
12 212.43
};
\addlegendentry{$L_{low}$=1}

\addplot[
    color=blue,
	fill,
]
table {
1   17.075	
2   34.57	
3   52.608	
4   71.315	
5   90.612	
6   110.39	
7   130.3	
8   149.75	
9   168.26	
10  185.32	
11  200.5	
12  213.59	
};
\addlegendentry{$L_{low}$=2}

\addplot[
    color=Black,
	fill,
]
table {
1  17.55
2  35.407
3  53.812
4  72.894
5  92.525
6  112.33
7  131.98
8  151.22
9  169.56
10 186.38
11 201.45
12 214.35
};
\addlegendentry{$L_{low}$=4}

\addplot[
    color=Gray,
	fill,
]
table {
1  17.92	
2  36.051	
3  54.747	
4  74.073	
5  93.607	
6  113.71	
7  133.52	
8  152.81	
9  170.94	
10 187.37	
11 202.29	
12 214.98	
};
\addlegendentry{$L_{low}$=8}

\addplot[
    color=Bittersweet,
	fill,
]
table {
1   18.297
2   36.933
3   55.527
4   74.908
5   95.1	
6   114.6
7   134.68
8   153.84
9   171.93
10  188.65
11  203.41
12  215.95
};
\addlegendentry{$L_{low}$=16}

\addplot[
color=BurntOrange,
	fill,
]
table {
1  18.769
2  37.824
3  56.324
4  75.769
5  96.056
6  116.3
7  136.13
8  154.83
9  172.57
10 188.71
11 203.12
12 216.07
};
\addlegendentry{$L_{low}$=32}

\end{axis}
\end{tikzpicture}
  \\
 \ref{LLRdist}
  \caption{LLR distribution for $N$=256, $R$=1/2, $C$=5, $E_b/N_0$=2.4 dB with different $L_{low}$.}
  \label{fig:LLRdist}
\end{figure}
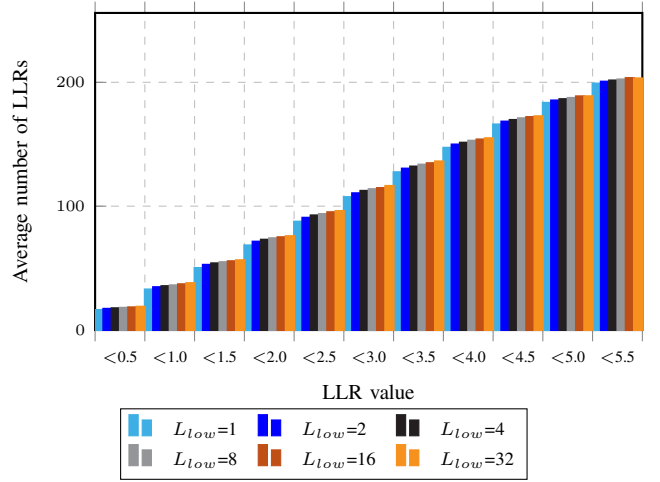

Consequently, given the unbalanced $L_{low}$ requirements, a decoder implementation with fixed list size performs a large amount of unnecessary operations and memory accesses, resulting in power consumption that is ultimately wasted. 
It would be advantageous to identify the required $L_{low}$ before the start of the decoding process, so that the list size can be reduced accordingly, effectively dividing the power consumption by up to a factor $L_{high}$. 
To attempt to do so, it is necessary to rely on the only information available before the decoding, i.e. the channel LLRs. 
Channel LLRs for an AWGN channel and binary phase shift keying (BPSK) modulation can be computed as $\boldsymbol{y}=2/\sigma^2 \cdot (1-2\boldsymbol{x} + E )$, where $\sigma$ is the standard deviation of the channel noise distribution, and the random variable $E\sim \mathcal{N}(0,\sigma^2)$ has equiprobable sign and magnitude that increases with decreasing probability. 
Large-magnitude LLRs are thus associated to correct bits with high probability, as it would require a very large $E$ (and thus a very improbable one) to move the symbol across the decision threshold (0 in BPSK) and increase its magnitude again. 
On the other hand, flipped bits are usually associated to small-magnitude LLRs, as smaller $E$ are more probable and can move the symbol across the decision threshold, but not increase its magnitude again. 
SC decoding uses this interpretation of the received LLR vector to correct errors. 
The operations used to propagate soft information through the SC decoding tree can however result in wrong decisions in case a high enough number of correct LLRs have smaller magnitude than LLRs whose bit has been flipped by channel noise (erroneous LLRs). 
This can happen when erroneous LLRs have large magnitudes, or when correct LLRs have small magnitudes. 
While the first case is rare and difficult to correct, the second case is common and can be corrected via SCL decoding, that is able to consider lower-probability decoding candidates. 

Fig. \ref{fig:LLRdist} shows the distribution of LLR values for $N=256$, $R=1/2$, $C=5$, averaged over $10^6$ transmitted codewords. 
The x axis shows a set of LLR magnitudes, and the y axis the average number of LLRs with magnitude lower than that. 
It can be seen that the LLRs are distributed differently depending on the $L_{low}$ required for correct decoding. 
When the required $L_{low}$ is small, a low number of LLRs have small magnitude: erroneous LLRs are mostly small, and correct LLRs tend to have larger magnitude. 
SC is well tuned for this situation, and a single decoder or a few parallel decoding paths are sufficient to correct the channel-induced errors. 
On the other hand, when the required $L_{low}$ is large, the LLRs are more concentrated around smaller magnitudes. 
In this situation many correct LLRs are small, and SC is more likely to make mistakes. 
Consequently, numerous parallel decoders are necessary to include low-probability paths among the final candidates.

Based on these results, two embodiments of input-distribution-aware SCL decoding (IDA-SCL) are proposed, a technique to dynamically reduce the active list size of SCL decoders before each decoding, that relies on simple observations on the channel LLRs. 
This technique can be combined with existing complexity- and latency reduction- methods \cite{PartSCL,ANNpredictor,GenFast,fastArdakani}. 

\subsection{Single-layer IDA-SCL}\label{subsec:SL}

\begin{figure}[t!]

    \centering
    \begin{minipage}{.25\textwidth}
        \centering
        		  \vspace{-12pt}
		  \begin{tikzpicture}
  \pgfplotsset{
    label style = {font=\fontsize{9pt}{7.2}\selectfont},
    tick label style = {font=\fontsize{7pt}{7.2}\selectfont}
  }

\begin{axis}[
	scale = 1,
	ymode=log,
	xlabel={$\phi$}, xlabel style={yshift=0.4em},
    ylabel={BLER}, ylabel style={yshift=-0.75em},
    ymajorgrids=true,
    xmajorgrids=true,
    grid style=dashed,
    mark options=solid,
    width=1\columnwidth, height=5.5cm,
    thick,
		xmin=0,
       xmax=20,
    mark size=3,
    legend style={
      anchor={center},
      cells={anchor=west},
      mark options=solid,
      column sep= 2mm,
      font=\fontsize{7pt}{7.2}\selectfont,
    },
    legend to name=BLER_param,
    legend columns=4,
]

\addplot[
    color=BurntOrange,
    thick,
    mark size=3,
]
table {
	0       0.00103297
	1        0.0010375
	2         0.001532
	3        0.0026918
	4        0.0055191
	5        0.0094823
	6         0.014203
	7         0.015411
	8         0.027345
	9         0.028082
	10        0.026089
	11        0.029172
	12        0.026302
	13        0.030075
	14        0.027601
	15        0.027005
	16        0.030675
	17        0.030303
	18         0.03775
	19        0.031746
	20        0.029718
};
\addlegendentry{$\gamma=0.5$}

\addplot[
    color=black,
    thick,
    mark size=3,
]
table {
	0          0.0010038	
	1         0.00093228	
	2         0.00097456	
	3         0.00095807	
	4         0.00080318	
	5          0.0010126	
	6         0.00096693	
	7          0.0016977	
	8          0.0021104	
	9          0.0029119	
	10         0.0043817	
	11          0.007326	
	12          0.010803	
	13          0.014198	
	14          0.018879	
	15          0.022563	
	16          0.023375	
	17          0.027996	
	18          0.029967	
	19          0.029019	
	20          0.032819	
};
\addlegendentry{$\gamma=1.0$}

\addplot[
    color=CornflowerBlue,
    thick,
    mark size=3,
]
table {
	0	    0.00082326
	1       0.00082608
	2       0.00086472
	3        0.0009342
	4       0.00081443
	5       0.00076376
	6         0.000809
	7       0.00084085
	8        0.0010154
	9       0.00092455
	10      0.00092386
	11       0.0011189
	12       0.0011257
	13       0.0015397
	14       0.0018294
	15       0.0022563
	16        0.003306
	17       0.0044849
	18       0.0063646
	19        0.010509
	20        0.012952
};
\addlegendentry{$\gamma=1.5$}

\addplot[
    color=OliveGreen,
    thick,
    mark size=3,
]
table {
	0	    0.00090128
	1       0.00080763
	2       0.00093981
	3       0.00082907
	4       0.00097923
	5       0.00097404
	6       0.00091093
	7       0.00097198
	8       0.00089451
	9       0.00085926
	10       0.0008391
	11      0.00087449
	12       0.0012136
	13      0.00093679
	14      0.00093693
	15      0.00078198
	16       0.0008152
	17      0.00098775
	18        0.001185
	19       0.0012725
	20       0.0014163
};
\addlegendentry{$\gamma=2.0$}


\end{axis}
\end{tikzpicture}
		        \end{minipage}%
    \begin{minipage}{0.25\textwidth}
        \centering
		  \begin{tikzpicture}
  \pgfplotsset{
    label style = {font=\fontsize{9pt}{7.2}\selectfont},
    tick label style = {font=\fontsize{7pt}{7.2}\selectfont}
  }

\begin{axis}[
	scale = 1,
	xlabel={$\phi$}, xlabel style={yshift=0.4em},
    ylabel={Complexity [\%] }, ylabel style={yshift=-0.75em},
    ymajorgrids=true,
    xmajorgrids=true,
    grid style=dashed,
    mark options=solid,
    width=1\columnwidth, height=5.5cm,
    thick,
       xmin=0,
       xmax=20,
    mark size=3,
    legend style={
      anchor={center},
      cells={anchor=west},
      mark options=solid,
      column sep= 2mm,
      font=\fontsize{7pt}{7.2}\selectfont,
    },
    legend to name=Comp_param,
    legend columns=2,
]

\addplot[
    color=BurntOrange,
    thick,
    mark size=3,
]
table {
	0	99.81
	1   98.68
	2   95.06
	3   87.16
	4   74.48
	5   59.34
	6   43.42
	7   30.19
	8   18.44
	9   12.27
	10  7.35
	11  4.93
	12  4.17
	13  3.53
	14  3.31
	15  3.20  
	16  3.13
	17  3.13
	18  3.13
	19  3.13
	20  3.13
};
\addlegendentry{$\gamma=0.5$}

\addplot[
    color=black,
    thick,
    mark size=3,
]
table {
	0		100.00
	1       100.00
	2       99.98
	3       99.90
	4       99.56
	5       98.90
	6       97.18
	7       94.10
	8       88.83
	9       81.22
	10      72.22
	11      60.89
	12      50.45
	13      39.33
	14      28.56
	15      19.76
	16      14.92
	17      9.91
	18      6.84
	19      5.06
	20      4.56
};
\addlegendentry{$\gamma=1.0$}

\addplot[
    color=CornflowerBlue,
    thick,
    mark size=3,
]
table {
	0	100.00
	1   100.00
	2   100.00
	3   100.00
	4   100.00
	5   99.99
	6   99.98
	7   99.93
	8   99.81
	9   99.53
	10  98.89
	11  97.71
	12  95.72
	13  92.57
	14  88.17
	15  82.09
	16  74.41
	17  66.03
	18  57.39
	19  47.00
	20  37.00
};
\addlegendentry{$\gamma=1.5$}

\addplot[
    color=OliveGreen,
    thick,
    mark size=3,
]
table {
	0	100.00
	1   100.00
	2   100.00
	3   100.00
	4   100.00
	5   100.00
	6   100.00
	7   100.00
	8   100.00
	9   100.00
	10  100.00
	11  99.98
	12  99.94
	13  99.86
	14  99.73
	15  99.39
	16  98.88
	17  97.84
	18  96.14
	19  93.98
	20  90.35
};
\addlegendentry{$\gamma=2.0$}


\end{axis}
\end{tikzpicture}
   \end{minipage}
    \ref{BLER_param}
    \caption{BLER and complexity for $N=128$, $R=1/2$, $C=2$, $E_b/N_0=3.25$, $L_{low}=1$, $L_{high}=32$ for different $\gamma$ and $\phi$.}
    \label{fig:param}
\end{figure}
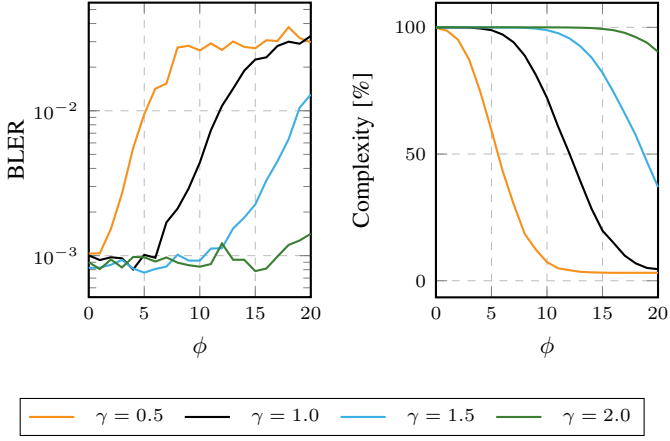

The first of the two proposed techniques is the single-layer IDA-SCL, as two thresholds $\gamma$ and $\phi$ are used to identify a single $L_{low}$ as an alternative to $L_{high}$. 
The channel LLRs that are lower or equal than $\gamma$ are counted; if their number is lower than $\phi$, then the decoding is attempted through SCL with $L_{low}$, otherwise $L_{high}$ is used. 
Given a combination of code parameters and $E_b/N_0$, the single-layer IDA-SCL requires the selection of $L_{low}$ and of both $\gamma$ and $\phi$, to minimize the decoding complexity in dependence of a target BLER. 

An immediate way to exploit the correlation between $L_{low}$ and the distribution of LLRs would be to set $\gamma$ and $\phi$ according to the average LLR distribution (e.g. Fig. \ref{fig:LLRdist}).
However, the average distribution represents a trend and, since the AWGN is a random process, strong variations in the LLR distribution are observed between realizations.
Consequently, a heuristic approach is necessary to choose $\gamma$ and $\phi$ values that allow to reliably select $L_{low}$ without causing undue BLER degradation.
To exemplify the threshold selection process, Fig. \ref{fig:param} plots the BLER and decoding complexity for $N=128$, $R=1/2$, $C=2$, $E_b/N_0=3.25$, $L_{low}=1$, $L_{high}=32$, at the variation of $\gamma$ and $\phi$. 
The complexity, intended here as the percentage of active paths with respect to the total, is computed as
\begin{equation} \label{eq:comp_SL}
100 \cdot \frac{\delta\cdot L_{low}+(1-\delta)\cdot L_{high}}{L_{high}}~,	
\end{equation}
where $\delta$ is the fraction of times that $L_{low}$ is chosen over $L_{high}$. 
The optimal combination of $\gamma$ and $\phi$ is the one that minimizes the complexity given a target maximum BLER. 
For instance, if the maximum acceptable BLER=$2\cdot10^{-3}$ in Fig. \ref{fig:param}, the following combinations of thresholds show better performance: for $\gamma=0.5$, $\phi<3$; for $\gamma=1.0$, $\phi<8$; for $\gamma=1.5$, $\phi<15$. 
The lowest complexity among these thresholds is 88.17\%, achieved with $\gamma=1.5$ and $\phi=14$. 
The difference between the SCL and IDA-SCL BLER expresses the misidentification rate, i.e. the normalized number of blocks where $L_{low}$ was selected and that resulted in a decoding error, while correct decoding could be achieved with $L_{high}$.

The threshold selection process is repeated for all the possible $L_{low}<L_{high}$, so that the $L_{low}$ that minimizes the complexity is selected to be used in single-layer IDA-SCL.

\subsection{Multi-layer IDA-SCL}\label{subsec:ML}

The single-layer IDA-SCL selects the best $L_{low}$ to minimize complexity. 
However, by considering multiple sets of $\gamma$ and $\phi$, each targeting a different $L_{low}$, it is possible to partially combine the complexity reduction effects of various single-layer IDA-SCL. 
In the multi-layer IDA-SCL, the decision conditions for increasing $L_{low}$ are checked sequentially until one is met, or until $L_{low}=L_{high}$, as shown in Algorithm \ref{alg:MLMESCL}.

\begin{algorithm}[t!]
\caption{Multi-layer IDA-SCL} \label{alg:MLMESCL}
\footnotesize
\begin{algorithmic}[1]
\FOR{$i = 0 \dots \log_2(L_{high})-1$}	
   \STATE $\texttt{LLR count}=0$
   \FOR{$j = 0 \dots N-1$}	
	   	\IF{$y[j]\le \gamma_i$}
		   \STATE $\texttt{LLR count}++$  
	\ENDIF 
   \ENDFOR   
   \IF{$\texttt{LLR count}<\phi_i$}
   	   \STATE $L_{low} = 2^i$	
	   \RETURN $\hat{\boldsymbol{u}} = \text{SCL decoding}(\boldsymbol{y}, L_{low})$ 
   \ENDIF		
\ENDFOR
\RETURN $\hat{\boldsymbol{u}} = \text{SCL decoding}(\boldsymbol{y}, L_{high})$
\end{algorithmic}
\end{algorithm}

As each threshold couple misidentifies a potentially different set of cases, the BLER degradation caused by each $L_{low}$ is partially accumulated, leading to unacceptable performance. 
For this reason, the optimal thresholds identified for each $L_{low}$ need to be decreased. 
Complexity is computed as
\begin{equation}\label{eq:comp_ML}
100 \cdot \sum_{i=0}^{\log_2L_{high}} \delta_i \cdot \frac{2^i}{L_{high}},
\end{equation}
where $\delta_i$ is the fraction of times that $L_{low}=2^i$ is selected.

\section{Results}\label{sec:results}

\begin{table*}[t!]
  \caption{$L_{low}$, $L_{high}$, $\gamma$ and $\phi$ for the observed codes at given $E_b/N_0$.}
	\centering
	\begin{tabular}{|c|c|c|c|c|c|c|}
		\hline
		 & \multicolumn{2}{c|}{$N=128$} & \multicolumn{2}{c|}{$N=256$} & \multicolumn{2}{c|}{$N=512$} \\
		\cline{2-7}
		    &	$R=1/2$ $C=2$ &	 $R=7/8$ $C=3$ & $R=1/2$ $C=5$ & $R=7/8$ $C=8$ & $R=1/2$ $C=8$ & $R=7/8$ $C=12$ \\
   		\hline
		$E_b/N_0$   &	3.25	&	5.30	&	2.40		& 4.85	&	2.60	& 4.45		\\
		$L_{high}$	&	16	&	8	&	32		&	32 &	32	&	32	\\
		$L_{low}$	&	8	&	4	&	16		&	16 &	16	&	16	\\
		$\gamma$	&	4.0	&	4.0	&  4.0		&	4.0 &	4.0	&	4.5	\\
		$\phi$		&	65	&	9	&	153		&	22 &	285	&	63	\\
		\hline	
	\end{tabular}
	\newline\newline
\label{tab:simparam}
\end{table*}

In this Section, simulation results for IDA-SCL are reported, considering polar codes constructed through density evolution with Gaussian approximation \cite{GA} with $\sigma=0.5$. Experiments have been run with the 5G standard sequence as well \cite{3GPP_R15}, and show that the effectiveness of IDA-SCL is independent from the code construction method.
The presented results concern the combinations of $N$, $R$ and $C$ already shown in Table \ref{tab:minL}.
The CRC polynomials used are 0x3 for $C=2$ and 3, 0x15 for $C=5$, 0xD5 for $C=8$, 0x80F for $C=12$. 
For each code parameter combination, Table \ref{tab:simparam} reports the selected $L_{high}$ (both single- and multi-layer IDA-SCL), the optimal $L_{low}$ for single-layer IDA-SCL, the $E_b/N_0$ at which the thresholds have been optimized (BLER$\approx10^{-3}$), and the value of $\gamma$ and $\phi$. 
The target maximum BLER has been chosen as the mid-point between the BLER of SCL with $L_{high}$ and with $L_{low}$. 
The majority of results have been obtained with $L_{high}=32$.
However, with $N=128$, $R=1/2$ there is very little difference between the BLER of SCL with $L=16$ and $L=32$; consequently, IDA-SCL uses $L_{high}=16$. 
In the same way, for $N=128$, $R=7/8$ the BLER of SCL with $L=8$, $L=16$, and $L=32$ is almost the same at the considered $E_b/N_0$. 
IDA-SCL thus uses $L_{high}=8$. 
In the figures, IDA-SCL with $L_{low}=a$ and $L_{high}=b$ is labeled as $L=a,b$.

\begin{figure}[t!]
  \centering
  \begin{tikzpicture}
  \pgfplotsset{
    label style = {font=\fontsize{9pt}{7.2}\selectfont},
    tick label style = {font=\fontsize{7pt}{7.2}\selectfont}
  }

\begin{axis}[
	scale = 1,
    ymode=log,
    xlabel={$E_b/N_0$ [\text{dB}]}, xlabel style={yshift=0.4em},
    ylabel={BLER}, ylabel style={yshift=-0.75em},
    grid=both,
    ymajorgrids=true,
    xmajorgrids=true,
    grid style=dashed,
    mark options=solid,
    width=1\columnwidth, height=5.5cm,
    thick,
        ymin=1e-4,
    mark size=3,
    legend style={
      anchor={center},
      cells={anchor=west},
      mark options=solid,
      column sep= 2mm,
      font=\fontsize{7pt}{7.2}\selectfont,
    },
    legend to name=BLER_0.5,
    legend columns=2,
]

\addplot[
    color=BurntOrange,
    mark=o,
    thick,
    mark size=3,
]
table {
2.5     0.00678887	
2.75    0.0037695	
3       0.0020734	
3.25    0.0010444	
3.5     0.00060047	
};
\addlegendentry{$N$=$128$ ME-SCL $L$=$8,16$}

\addplot[
    color=BurntOrange,
    mark=o,
    thick,
	dashed,
    mark size=3,
]
table {
2.5     0.00922084	 
2.75    0.0047328
3       0.0024997	 
3.25    0.0011663
3.5     0.00057514	 
};
\addlegendentry{$N$=$128$ SCL $L$=$8$}

\addplot[
    color=BurntOrange,
    mark=o,
    thick,
	dotted,
    mark size=3,
]
table {
2.5     0.00673764	 
2.75    0.00377271
3       0.0019323	 
3.25    0.000922891
3.5     0.00054512	 
};
\addlegendentry{$N$=$128$ SCL $L$=$16$}

\addplot[
    color=CornflowerBlue,
    mark=x,
    thick,
    mark size=3,
]
table {
1.65       0.0231911	
1.9        0.010733	    
2.15       0.0042885	
2.4        0.0014933	
2.65       0.00051324	
};
\addlegendentry{$N$=$256$ ME-SCL $L$=$16,32$}

\addplot[
    color=CornflowerBlue,
    mark=x,
    thick,
	dashed,
    mark size=3,
]
table {
1.65       0.0404694	
1.9        0.016537	   
2.15       0.0061323	
2.4       0.0022686	   
2.65      0.00059485   
};
\addlegendentry{$N$=$256$ SCL $L$=$16$}

\addplot[
    color=CornflowerBlue,
    mark=x,
    thick,
	dotted,
    mark size=3,
]
table {
1.65       0.0224266
1.9         0.01085	
2.15       0.0039502
2.4       0.0011783	
2.65      0.00038644
};
\addlegendentry{$N$=$256$ SCL $L$=$32$}

\addplot[
    color=black,
    mark=square,
    thick,
    mark size=3,
]
table {
1.7       0.0923361	
2          0.026745	
2.3       0.0061728	
2.6       0.0015488	
2.9      0.00031206	
};
\addlegendentry{$N$=$512$ ME-SCL $L$=$16,32$}

\addplot[
    color=black,
    mark=square,
    thick,
	dashed,
    mark size=3,
]
table {
1.7           0.124	
2          0.044964	
2.3        0.012545	
2.6       0.0026544	
2.9      0.00036593	
};
\addlegendentry{$N$=$512$ SCL $L$=$16$}

\addplot[
    color=black,
    mark=square,
    thick,
	dotted,
    mark size=3,
]
table {
1.7       0.0923361
2          0.026745	
2.3        0.006152
2.6       0.0011342
2.9      0.00013823
};
\addlegendentry{$N$=$512$ SCL $L$=$32$}

\end{axis}
\end{tikzpicture}
  \\
 \ref{BLER_0.5}
  \caption{BLER with SCL and single-layer IDA-SCL for $R$=$1/2$.}
  \label{fig:BLER05}
\end{figure}
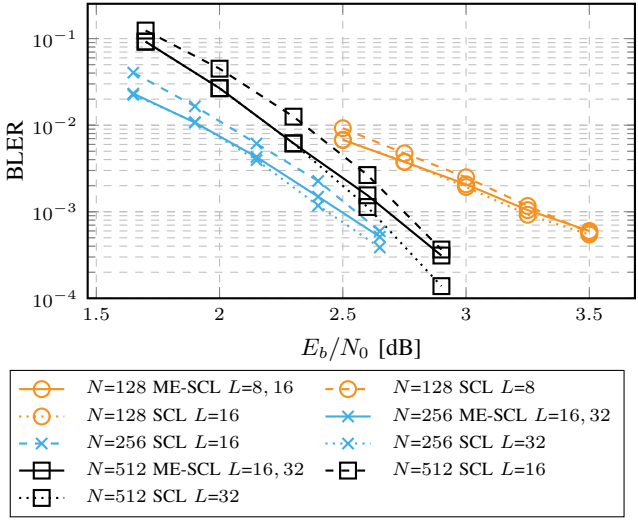

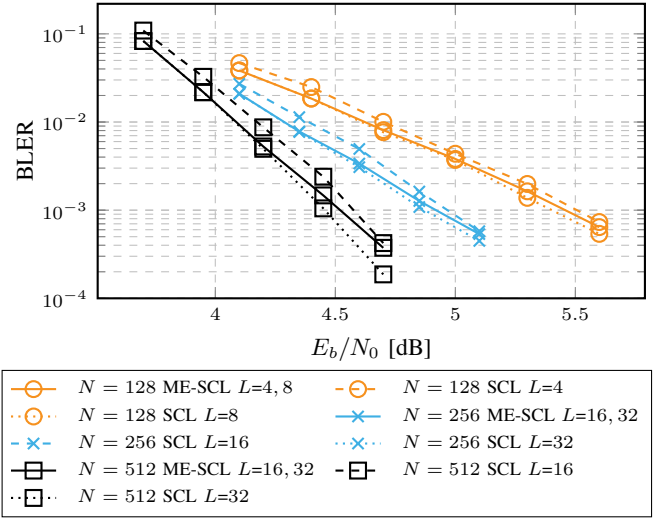
\begin{figure}[t!]
  \centering
  \begin{tikzpicture}
  \pgfplotsset{
    label style = {font=\fontsize{9pt}{7.2}\selectfont},
    tick label style = {font=\fontsize{7pt}{7.2}\selectfont}
  }

\begin{axis}[
	scale = 1,
    ymode=log,
    xlabel={$E_b/N_0$ [\text{dB}]}, xlabel style={yshift=0.4em},
    ylabel={BLER}, ylabel style={yshift=-0.75em},
    grid=both,
    ymajorgrids=true,
    xmajorgrids=true,
    grid style=dashed,
    mark options=solid,
    width=1\columnwidth, height=5.5cm,
    thick,
        ymin=1e-4,
    mark size=3,
    legend style={
      anchor={center},
      cells={anchor=west},
      mark options=solid,
      column sep= 2mm,
      font=\fontsize{7pt}{7.2}\selectfont,
    },
    legend to name=BLER_0.875,
    legend columns=2,
]

\addplot[
    color=BurntOrange,
    mark=o,
    thick,
    mark size=3,
]
table {
4.1       0.0384172
4.4        0.018471
4.7       0.0080893
5       0.0038123	
5.3       0.0016369
5.6      0.00064235
};
\addlegendentry{$N=128$ ME-SCL $L$=$4,8$}

\addplot[
    color=BurntOrange,
    mark=o,
    thick,
	dashed,
    mark size=3,
]
table {
 4.1         0.04662	
 4.4        0.024931	
 4.7        0.010047	
 5       0.0043647	
 5.3       0.0019832	
 5.6      0.00073382	
};
\addlegendentry{$N=128$ SCL $L$=$4$}

\addplot[
    color=BurntOrange,
    mark=o,
    thick,
	dotted,
    mark size=3,
]
table {
4.1       0.0384172
4.4        0.018471
4.7       0.0076599
5       0.0037121	
5.3       0.0013794
5.6      0.00053858 
};
\addlegendentry{$N=128$ SCL $L$=$8$}

\addplot[
    color=CornflowerBlue,
    mark=x,
    thick,
    mark size=3,
]
table {
4.1        0.020938	 
4.35       0.0078009
4.6        0.0033786
4.85       0.0012713
5.1        0.00053576
};
\addlegendentry{$N=256$ ME-SCL $L$=$16,32$}

\addplot[
    color=CornflowerBlue,
    mark=x,
    thick,
	dashed,
    mark size=3,
]
table {
4.1       0.0271003	
4.35        0.011368
4.6       0.0049451	
4.85       0.0016273
5.1      0.00057682	
};
\addlegendentry{$N=256$ SCL $L$=$16$}

\addplot[
    color=CornflowerBlue,
    mark=x,
    thick,
	dotted,
    mark size=3,
]
table {
4.1        0.020938	
4.35       0.0077012
4.6       0.0030749	
4.85       0.0010843
5.1      0.00044828	
};
\addlegendentry{$N=256$ SCL $L$=$32$}

\addplot[
    color=black,
    mark=square,
    thick,
    mark size=3,
]
table {
3.7       0.0829876	 
3.95        0.021801
4.2       0.0052765	 
4.45       0.0014712
4.7      0.00037395	 
};
\addlegendentry{$N=512$ ME-SCL $L$=$16,32$}

\addplot[
    color=black,
    mark=square,
    thick,
	dashed,
    mark size=3,
]
table {
3.7           0.109	 
3.95        0.032552
4.2        0.008704	 
4.45       0.0023964
4.7       0.0004233	 
};
\addlegendentry{$N=512$ SCL $L$=$16$}

\addplot[
    color=black,
    mark=square,
    thick,
	dotted,
    mark size=3,
]
table {
3.7       0.0829876	 
3.95        0.021801
4.2       0.0049687	 
4.45       0.0010482
4.7      0.00018692	 
};
\addlegendentry{$N=512$ SCL $L$=$32$}

\end{axis}
\end{tikzpicture}
  \\
 \ref{BLER_0.875}
  \caption{BLER with SCL and single-layer IDA-SCL for $R$=$7/8$.}
  \label{fig:BLER0875}
\end{figure}

\begin{figure}[t!]
  \centering
  \begin{tikzpicture}
  \pgfplotsset{
    label style = {font=\fontsize{9pt}{7.2}\selectfont},
    tick label style = {font=\fontsize{7pt}{7.2}\selectfont}
  }

\begin{axis}[
	scale = 1,
    xlabel={$L_{low}$}, xlabel style={yshift=0.4em},
    ylabel={Complexity [\%] }, ylabel style={yshift=-0.75em},
    ymajorgrids=true,
    xmajorgrids=true,
    grid style=dashed,
    mark options=solid,
    width=1\columnwidth, height=5.5cm,
    thick,
    mark size=3,
	xtick={1,2,3,4,5,6},
    xticklabels={$1$,$2$,$4$,$8$,$16$,$32$},
    legend style={
      anchor={center},
      cells={anchor=west},
      mark options=solid,
      column sep= 2mm,
      font=\fontsize{7pt}{7.2}\selectfont,
    },
    legend to name=Comp_evo1,
    legend columns=2,
]

\addplot[
    color=BurntOrange,
    mark=o,
    thick,
    mark size=3,
]
table {
1  96.1
2  85.5
3  59.0
4  57.7
5 100.0
};
\addlegendentry{$N$=$128$ $R$=$1/2$ $E_b/N_0$=$3.25$}

\addplot[
    color=BurntOrange,
    mark=o,
	dashed,
    thick,
    mark size=3,
]
table {
1  82.7
2  72.5
3  56.8
4  100.0
};
\addlegendentry{$N$=$128$ $R$=$7/8$ $E_b/N_0$=$5.30$}

\addplot[
    color=CornflowerBlue,
    mark=x,
    thick,
    mark size=3,
]
table {
1  99.2
2  95.5
3  84.8
4  63.9
5 61.3
6 100.0
};
\addlegendentry{$N$=$256$ $R$=$1/2$ $E_b/N_0$=$2.40$}

\addplot[
    color=CornflowerBlue,
    mark=x,
	dashed,
    thick,
    mark size=3,
]
table {
1 98.3
2 91.8
3 83.9
4 73.9
5 59.1
6 100.0
};
\addlegendentry{$N$=$256$ $R$=$7/8$ $E_b/N_0$=$4.85$}

\addplot[
    color=black,
    mark=square,
    thick,
    mark size=3,
]
table {
1 98.9 
2 97.9 
3 89.7 
4 77.1 
5 69.0
6 100.0
};
\addlegendentry{$N$=$512$ $R$=$1/2$ $E_b/N_0$=$2.6$}

\addplot[
    color=black,
    mark=square,
	dashed,
    thick,
    mark size=3,
]
table {
1  98.8
2  95.7
3  85.8
4  80.4
5 64.7
6 100.0
};
\addlegendentry{$N$=$512$ $R$=$7/8$ $E_b/N_0$=$4.45$}


\end{axis}
\end{tikzpicture}
  \\
 \ref{Comp_evo1}
  \caption{Single-layer IDA-SCL complexity with different $L_{low}$.}
  \label{fig:comp1}
\end{figure}
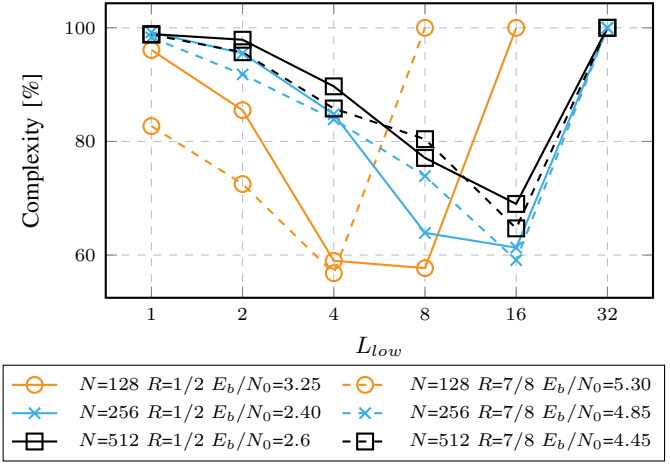

Fig. \ref{fig:BLER05}-\ref{fig:BLER0875} plot the BLER for the three considered code lengths, for $R=1/2$ and $R=7/8$, respectively. 
Both single-layer IDA-SCL and standard SCL are portrayed.
It can be observed that the BLER of IDA-SCL follows closely that of SCL with $L_{high}$, starting to diverge when closer to the $E_b/N_0$ for which the thresholds were optimized, and converges to the BLER of SCL with $L_{low}$ at higher $E_b/N_0$.
At lower $E_b/N_0$, channel LLRs have smaller magnitude: $\phi$ tends to be very restrictive, and a higher percentage of vectors is decoded with $L_{high}$. 
In the same way, at higher $E_b/N_0$ fewer LLRs are below the $\gamma$ threshold, and $L_{low}$ is prioritized; since thresholds have to be optimized for a given $E_b/N_0$, the decoder incurs a higher misidentification rate.

\begin{figure}[t!]
  \centering
  \begin{tikzpicture}
  \pgfplotsset{
    label style = {font=\fontsize{9pt}{7.2}\selectfont},
    tick label style = {font=\fontsize{7pt}{7.2}\selectfont}
  }

\begin{axis}[
	scale = 1,
    xlabel={$E_b/N_0$ [\text{dB}]}, xlabel style={yshift=0.4em},
    ylabel={Complexity [\%] }, ylabel style={yshift=-0.75em},
    grid=both,
    ymajorgrids=true,
    xmajorgrids=true,
    grid style=dashed,
    mark options=solid,
    width=1\columnwidth, height=5.5cm,
    thick,
    mark size=3,
    legend style={
      anchor={center},
      cells={anchor=west},
      mark options=solid,
      column sep= 2mm,
      font=\fontsize{7pt}{7.2}\selectfont,
    },
    legend to name=Comp_evo2,
    legend columns=2,
]

\addplot[
    color=BurntOrange,
    mark=o,
    thick,
    mark size=3,
]
table {
2.5    94.4265
2.75   83.776
3      69.311
3.25   57.738
3.5    51.9492
};
\addlegendentry{$N$=128 $R$=1/2 $L$=$8,16$}

\addplot[
    color=BurntOrange,
    mark=o,
	dashed,
    thick,
    mark size=3,
]
table {
4.1    97.40715
4.4    91.1525
4.7    79.4855
5      66.8655
5.3    56.8425
5.6    52.0887
};
\addlegendentry{$N$=$128$ $R$=$7/8$ $L$=$4,8$}

\addplot[
    color=CornflowerBlue,
    mark=x,
    thick,
    mark size=3,
]
table {
1.65      99.55963
1.9       95.16525
2.15      80.6075
2.4       61.2685
2.65      51.881
};
\addlegendentry{$N$=256 $R$=1/2 $L$=$16,32$}

\addplot[
    color=CornflowerBlue,
    mark=x,
	dashed,
    thick,
    mark size=3,
]
table {
4.1     97.03775
4.35    88.369
4.6     73.0825
4.85    59.0585
5.1     52.1714
};
\addlegendentry{$N$=$256$ $R$=$7/8$ $L$=$16,32$}

\addplot[
    color=black,
    mark=square,
    thick,
    mark size=3,
]
table {
1.7  100
2    99.94649
2.3  96.23445
2.6  69.036
2.9  50.9701
};
\addlegendentry{$N$=$512$ $R$=$1/2$ $L$=$16,32$}

\addplot[
    color=black,
    mark=square,
	dashed,
    thick,
    mark size=3,
]
table {
3.7     100
3.95    99.05135
4.2     88.7585
4.45    64.653
4.7     51.6825
};
\addlegendentry{$N$=$512$ $R$=$7/8$ $L$=$16,32$}

\end{axis}
\end{tikzpicture}
  \\
 \ref{Comp_evo2}
  \caption{Single-layer IDA-SCL complexity at different $E_b/N_0$.}
  \label{fig:comp2}
\end{figure}
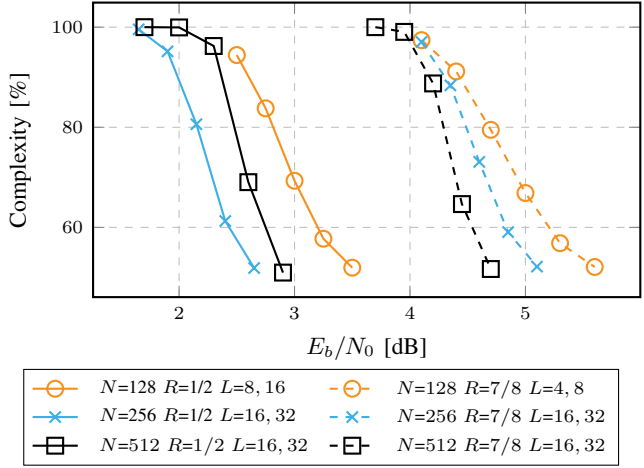

Fig. \ref{fig:comp1} portrays the complexity percentage for the observed codes at the threshold optimization $E_b/N_0$, with single-layer IDA-SCL. 
Each point in the graph uses the optimal thresholds for the associated $L_{low}$. 
As noted earlier in this section, for $N=128$, $R=1/2$, the selected $L_{high}=16$; thus, $L_{low}=16$ constitutes the $100\%$ complexity mark. 
Similarly, $L_{low}=8$ represents $100\%$ complexity in case of $N=128$, $R=7/8$. 
It can be seen that in all observed cases, $L_{low}=L_{high}/2$ achieves the lowest decoding complexity.
As the gap between $L_{high}$ and $L_{low}$ increases, the chances of misidentification rise as well, and stricter thresholds need to be used to remain within the target BLER. 
This leads to a reduced complexity gain, that more than counterbalances the positive effect of the smaller $L_{low}$ in (\ref{eq:comp_SL}). 
Fig. \ref{fig:comp2} depicts the evolution of the complexity of the observed codes as the channel conditions change, using the parameter set in Table \ref{tab:simparam}.
To minimize complexity for all channel conditions, a different threshold set should be used at each $E_b/N_0$; a single optimization point is shown nevertheless to bring substantial complexity reduction also before the working point, and thus at a lower BLER degradation cost (see Fig. \ref{fig:BLER05}-\ref{fig:BLER0875}).
Complexity keeps decreasing as the $E_b/N_0$ increases, reaching a minimum of 50\% since $L_{low}=L_{high}/2$, but at a higher BLER.

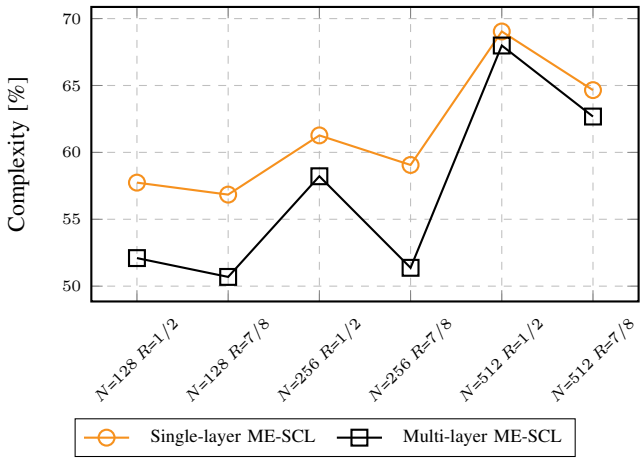
\begin{figure}[t!]
  \centering
  \begin{tikzpicture}
  \pgfplotsset{
    label style = {font=\fontsize{9pt}{7.2}\selectfont},
    tick label style = {font=\fontsize{6pt}{7.2}\selectfont}
  }

\begin{axis}[
	scale = 1,
    ylabel={Complexity [\%] }, ylabel style={yshift=-0.75em},
    ymajorgrids=true,
    xmajorgrids=true,
    grid style=dashed,
    mark options=solid,
    width=1\columnwidth, height=5.5cm,
    thick,
    mark size=3,
	xtick={1,2,3,4,5,6},
    xticklabels={$N$=$128$ $R$=$1/2$,$N$=$128$ $R$=$7/8$,$N$=$256$ $R$=$1/2$,$N$=$256$ $R$=$7/8$,$N$=$512$ $R$=$1/2$, $N$=$512$ $R$=$7/8$},
	xticklabel style={rotate=45},
    legend style={
      anchor={center},
      cells={anchor=west},
      mark options=solid,
      column sep= 2mm,
      font=\fontsize{7pt}{7.2}\selectfont,
    },
    legend to name=Comp_ML,
    legend columns=2,
]

\addplot[
    color=BurntOrange,
    mark=o,
    thick,
    mark size=3,
]
table {
1 57.738 
2 56.8425 
3 61.2685 
4 59.0585 
5 69.036
6 64.653
};
\addlegendentry{Single-layer ME-SCL}

\addplot[
    color=black,
    mark=square,
    thick,
    mark size=3,
]
table {
1 52.098
2 50.690 
3 58.2149 
4 51.3708
5 67.987
6 62.678
};
\addlegendentry{Multi-layer ME-SCL}


\end{axis}
\end{tikzpicture}
  \\
 \ref{Comp_ML}
  \caption{Multi-layer and single-layer IDA-SCL complexity.}
  \label{fig:comp_multi}
\end{figure}

Multi-layer IDA-SCL combines the complexity reduction potential of multiple single-layer thresholds.
As detailed in Section \ref{subsec:ML}, the thresholds for each layer have been decreased so that the BLER of multi-layer IDA-SCL matches that of single-layer IDA-SCL shown in Fig. \ref{fig:BLER05}-\ref{fig:BLER0875}. 
In Fig. \ref{fig:comp_multi}, the complexity of multi-layer and single-layer IDA-SCL are then compared under the conditions depicted in Table \ref{tab:simparam}. 
Multi-layer IDA-SCL brings 1\%-8\% additional decoding complexity reduction with respect to single-layer IDA-SCL.

IDA-SCL fills a currently empty niche in the landscape of on-the-fly power-reduction techniques.
IDA-SCL sacrifices some BLER to save complexity without incurring additional latency, whereas ASCL \cite{ASCL} imposes additional latency to reduce complexity at no BLER cost.
If the system can accept a decoding latency equal to multiple subsequent decoding attempts, ASCL can take full advantage of its sequential decoding features, and results in more substantial power saving than IDA-SCL at no BLER cost. 
On the other hand, if the system can accept only a latency equal to a single decoding attempt, ASCL inherently reverts to SCL. 
As ASCL foresees no way to choose between one list size and the other without observing the decoding outcome, it either results in no power saving ($L=L_{high}$) or in unacceptable BLER ($L<L_{high}$), while IDA-SCL can provide substantial complexity reduction at an acceptable BLER cost and no additional latency.


\section{Conclusion}\label{sec:conc}

In this work I have introduced input-distribution-aware SCL (IDA-SCL) decoding of polar codes, a technique to select the list size of SCL decoders by observing the distribution of channel LLRs. 
It can be used as a power-reduction technique in hardware SCL decoders, allowing to dynamically decrease the list size before each decoding, and thus deactivating part of the internal parallelism. 
IDA-SCL is based on simple, implementation-friendly threshold comparisons, and can be combined with existing complexity- and latency-reduction techniques. 
Two embodiments have been proposed, and shown to reduce the decoding complexity of SCL decoding of almost 50\% while meeting an error-correction performance target. 


\begin{thebibliography}{10}

\bibitem{ArikanFirst}
E.~Arikan,
\newblock ``Channel polarization: A method for constructing capacity-achieving
  codes for symmetric binary-input memoryless channels,''
\newblock {\em IEEE Transactions on Information Theory}, vol. 55, no. 7, pp.
  3051--3073, July 2009.

\bibitem{3GPP_R15}
$3^{\text{rd}}$ Generation Partnership Project~({3GPP}),
\newblock ``Multiplexing and channel coding,''
\newblock {\em 3GPP 38.212 V.15.3.0}, 2018.

\bibitem{Mehmood:20}
T.~Mehmood, M.~P. Yankov, A.~Fisker, K.~Gormsen, and S.~Forchhammer,
\newblock ``Rate-adaptive concatenated polar-staircase codes for data center
  interconnects,''
\newblock in {\em Optical Fiber Communication Conference (OFC) 2020}. 2020, p.
  Th1I.6, Optical Society of America.

\bibitem{PPCjournal}
C.~{Condo}, V.~{Bioglio}, H.~{Hafermann}, and I.~{Land},
\newblock ``Practical product code construction of polar codes,''
\newblock {\em IEEE Transactions on Signal Processing}, vol. 68, pp.
  2004--2014, 2020.

\bibitem{TalSCL}
I.~{Tal} and A.~{Vardy},
\newblock ``List decoding of polar codes,''
\newblock {\em IEEE Transactions on Information Theory}, vol. 61, no. 5, pp.
  2213--2226, May 2015.

\bibitem{PartSCL}
S.~A. {Hashemi}, A.~{Balatsoukas-Stimming}, P.~{Giard}, C.~{Thibeault}, and
  W.~J. {Gross},
\newblock ``Partitioned successive-cancellation list decoding of polar codes,''
\newblock in {\em 2016 IEEE International Conference on Acoustics, Speech and
  Signal Processing (ICASSP)}, 2016, pp. 957--960.

\bibitem{PartSCL_TCOM}
S.~A. {Hashemi}, M.~{Mondelli}, S.~H. {Hassani}, C.~{Condo}, R.~L. {Urbanke},
  and W.~J. {Gross},
\newblock ``Decoder partitioning: Towards practical list decoding of polar
  codes,''
\newblock {\em IEEE Transactions on Communications}, vol. 66, no. 9, pp.
  3749--3759, 2018.

\bibitem{ANNpredictor}
W.~{Song}, Y.~{Fu}, Q.~{Chen}, L.~{Li}, and C.~{Zhang},
\newblock ``{ANN} based adaptive successive cancellation list decoder for polar
  codes,''
\newblock in {\em 2019 IEEE 13th International Conference on ASIC (ASICON)},
  2019, pp. 1--4.

\bibitem{ASCL}
B.~{Li}, H.~{Shen}, and D.~{Tse},
\newblock ``An adaptive successive cancellation list decoder for polar codes
  with cyclic redundancy check,''
\newblock {\em IEEE Communications Letters}, vol. 16, no. 12, pp. 2044--2047,
  2012.

\bibitem{CA_SCL}
K.~{Niu} and K.~{Chen},
\newblock ``{CRC}-aided decoding of polar codes,''
\newblock {\em IEEE Communications Letters}, vol. 16, no. 10, pp. 1668--1671,
  October 2012.

\bibitem{GenFast}
C.~{Condo}, V.~{Bioglio}, and I.~{Land},
\newblock ``Generalized fast decoding of polar codes,''
\newblock in {\em 2018 IEEE Global Communications Conference (GLOBECOM)}, 2018,
  pp. 1--6.

\bibitem{fastArdakani}
M.~H. {Ardakani}, M.~{Hanif}, M.~{Ardakani}, and C.~{Tellambura},
\newblock ``Fast successive-cancellation-based decoders of polar codes,''
\newblock {\em IEEE Transactions on Communications}, vol. 67, no. 7, pp.
  4562--4574, 2019.

\bibitem{GA}
P.~{Trifonov},
\newblock ``Efficient design and decoding of polar codes,''
\newblock {\em IEEE Transactions on Communications}, vol. 60, no. 11, pp.
  3221--3227, 2012.

\end{thebibliography}

\end{document}